\documentclass[reprint, showpacs, superscriptaddress]{revtex4-1}
\usepackage[utf8]{inputenc}
\usepackage[english]{babel}
\usepackage{amsmath}
\usepackage{hyperref}
\usepackage{amsfonts}
\usepackage{amssymb}
\usepackage{graphicx}
\usepackage{graphics}
\usepackage{multirow}
\usepackage{natbib}
\usepackage{setspace}
\usepackage{hhline}
\usepackage[left=2cm,right=2cm,top=2cm,bottom=2cm]{geometry}

\begin{document}
\title{Weak quadrupole moments}

\author{B.G.C. Lackenby}
\affiliation{School of Physics, University of New South Wales,  Sydney 2052,  Australia}
\author{V.V. Flambaum}
\affiliation{School of Physics, University of New South Wales,  Sydney 2052,  Australia}
\affiliation{Johannes Gutenberg-Universit\"at Mainz, 55099 Mainz, Germany}
\begin{abstract}
We introduce the weak quadrupole moment of nuclei, related to the quadrupole distribution of the weak charge in the nucleus.  The weak quadrupole moment produces a tensor weak interaction  between the nucleus and electrons and can be observed in atomic and molecular experiments measuring parity nonconservation. The dominating contribution to the weak quadrupole is given by the quadrupole moment of the neutron distribution, therefore, corresponding experiments should allow one to measure the neutron quadrupoles. Using the deformed oscillator model and the Schmidt model we calculate the quadrupole distributions of neutrons, $Q_{n}$, the weak quadrupole moments, $Q_{W}^{(2)}$, and the Lorentz Invariance violating energy shifts in $^{9}$Be, $^{21}$Ne , $^{27}$Al, $^{131}$Xe, $^{133}$Cs, $^{151}$Eu, $^{153}$Eu, $^{163}$Dy, $^{167}$Er, $^{173}$Yb, $^{177}$Hf, $^{179}$Hf, $^{181}$Ta, $^{201}$Hg and $^{229}$Th. 
\end{abstract}

\maketitle
Non-spherical nuclei present a lucrative avenue for studying the existence and magnitude of second order tensor properties due to the collective properties of deformed nuclei. In this work we focus on the quadrupole moment of the nonspherical  distribution of neutrons in the nucleus, $Q_{n}$,  the weak quadrupole moment (WQM), $Q_{W}^{(2)}$,  and the violation of Local Lorentz invariance (LLI) in the nucleon sector. These properties have yet to be experimentally detected though there are constraints for violation of LLI. In this work we will show how these properties are enhanced in deformed nuclei and therefore present a new possibility for  measurement or developing further constraints on their existence.  In Section \ref{sec:Deformed} we calculate the the neutron quadrupole moment of the nucleus (NQMN)  and local Lorentz invariance violation (LLIV) in deformed nuclei using the Nilsson model. In Section \ref{sec:Minimal} we discuss NQMN and LLIV in nuclei with a small  deformation and in Section \ref{sec:PNC} we discuss the NQMN and WQM in parity nonconserving (PNC) effects. 

Measuring the NQMN, $Q_{n}$,  is a very difficult  task as neutrons are electrically neutral particles unlike the electric quadrupole moment of the nucleus which was first observed and measured nearly a century ago by H. Schuler and Th. Schmidt \cite{Schuler1935(1), Schuler1935(2), Casimir1935} by studying the hyperfine structure of rare earth elements. The NQMN is an important property which will give insight not only into the structure of  atomic nuclei but also other dense collections of neutrons such as neutron stars \cite{Brown2000, Furnstahl2002, Typel2001, Reinhard2010} and also the theory of atomic parity nonconservation. For the last two decades there has been increasing interest in understanding the distribution of neutrons compared to protons in atomic nuclei known as the neutron skin. This focus has largely been on the spherical distribution of neutrons (measurement of the root mean square radius of neutrons) and there has been a large amount of experimental effort in measuring the spherical distribution \cite{Clark2003, Trzcinska2001, Lenske2009, Abrahamyan2012}. 

The PNC effects appear due to mixing of opposite parity states in atoms and molecules by the weak interaction between the nucleus and electrons. The field  of PNC in atoms and molecules has been thoroughly reviewed in Refs.  \cite{KhriplovichPNC,GingesReview,RobertsReview}. It was noted in Ref.  \cite{FS78} that the nuclear quadrupole moment induces a tensor PNC weak interaction between the nucleus and electrons (see also  \cite{KhriplovichPNC,KP91}). In Ref. \cite{Flambaum2016} it was argued that these tensor  effects of the weak quadrupole moments are strongly enhanced for deformed nuclei and may get a significant additional enhancement due to mixing of  close atomic and molecular levels of opposite parity with a  difference of the electron angular momenta $|J_1-J_2| \le 2$. These selection rules are similar to that for the effects of the time reversal ($T$) and parity ($P$) violating nuclear magnetic quadrupole moment (MQM). Therefore, nuclei, molecules and molecular levels  suggested for the MQM search in Ref. \cite{Flambaum2014},  for example,  $|\Omega |=1$ doublets in the molecules $^{177}$HfF+, $^{229}$ThO, $^{181}$TaN will also have enhanced effects of the weak quadrupole (see section \ref{sec:PNC}). Following our proposal, there has been recent experimental interest in using the weak quadrupole moment to study PNC effects in HfF$^{+}$ molecules (used to measure electron electric dipole moment \cite{Cairncross2017}) and parity violation in $^{173}$Yb \cite{Antypas2017} and $^{163}$Dy \cite{Leefer2017} atoms. 

The other second rank tensor property we study is related to violation of Local Lorentz invariance (LLI). The physical property of LLI is fundamental in all disciplines of physics. It states that the laws of physics are invariant under transformations of velocities and orientations (boosts and rotations). However within the study of unification theories the possibility of local Lorentz invariance violation (LLIV) has been suggested and studied in a variety of experiments. These searches have been motivated by suggestions of LLIV in high energy theories such as string theory \cite{Kostelecky1995, Mavromatos2007,  Pospelov2012, Kostelecky1989, Liberati2013} which occur above the currently unaccessible Plank energy, $M_P$. It is expected  that signatures of these high energy phenomena will appear heavily suppressed in low energy electroweak experiments ($m_{ew}$)  to an order of $m_{ew}/M_{P} \approx \times 10^{-17}$\cite{Kostelecky1995}. This suggests that high energy phenomena such as LLIV can be observed and studied in a low energy regime with exceptionally sensitive measurements in nuclear and atomic systems. The seminal work on the systematic study of LLIV  was performed in \cite{Colladay1998} and \cite{Kostelecky1999} where a general form of Standard Model Extension (SME) Lagrangian  has been presented to include both $CPT$-odd and $CPT$-even LLIV terms  to create a unified model.\

Attempts to measure possible violations of LLI date back a century with the classic experiment performed by Michelson and Morley \cite{Michelson1887}. Recent attempts to detect LLIV (in the matter sector specifically) are more sophisticated using a variety of techniques such as resonance cavities \cite{Muller2005, Muller2003, Wolf2004} and Doppler shift experiments \cite{Lane2005, Saathoff2003}. However, the best current laboratory limits are from clock comparison experiments \cite{Prestage1985, Chupp1989, Hohensee2013, Dzuba2016} in particular the tensor LLIV parameters $c_{ij}$ for the neutron in \cite{Smiciklas2011} using a $^{21}$Ne clock and for the proton in \cite{Wolf2006} using $^{133}$Cs clocks which was further constrained by 4 orders of magnitude in \cite{Flambaum2016}.  In Ref. \cite{Brown2016} the $s-d$ shell model calculations of the tensor LLIV effects have been performed in $^{21}$Ne, $^{131}$Xe and $^{201}$Hg nuclei. It was demonstrated that virtual excitations from the nuclear core enhance the nuclear quadrupole moment and suppress the LLIV tensor.  The results of the $s-d$ shell model calculations have been supported by the calculations in the Hartree-Fock-Bogoliubov model published in the same paper \cite{Brown2016}.

The best laboratory limits on the Michelson-Morley type LLIV tensor in the photon sector has been obtained in Ref. \cite{FlambaumRomalis2017}  using the $^{21}$Ne clock data from  \cite{Smiciklas2011}.  
The limits on the LLIV parameters  obtained using astrophysical measurements and laboratory experiments can be found in \cite{Kostelecky1999, LorentzDataTables2017}. In the present paper we study  the LLIV momentum tensor Hamiltonian $\delta H$  \cite{Kostelecky1999}
\begin{align} \label{eq:HamilQuadShift}
\delta H = \left[-c_{ij} -\dfrac{1}{2}c_{00}\delta_{ij}\right]\dfrac{p_ip_j}{m^2},
\end{align}
 where $m$ is the nucleon mass,  to describe the tensor parameter $c_{ij}$  for the LLIV interaction \cite{Kostelecky1999} in the laboratory frame. In this paper we will show that there exists a collective effect of the momentum tensor in deformed nuclei and show that certain deformed nuclei of experimental interest present a promising option for constraining LLIV even further. \\ 

\section{Deformed Nuclei \label{sec:Deformed}}

In this paper we will use the empirically successful Nilsson model \cite{Nilsson1955, BohrMottVol2} (deformed oscillator model)  to describe the single particle model states of the constituent nucleons in deformed nuclei and calculate the magnitude of the quadrupole and momentum tensors. The Nilsson Hamiltonian which governs the system is,
\begin{align} \label{eq:NilssonHamiltonian}
H_{Nilsson} &= -\dfrac{\hbar^2}{2m}\Delta +\dfrac{m}{2}\left( \omega_z^2 z^2 + \omega_{\perp}^2\left( x^2 +  y^2\right) \right)
\end{align}
which has the form of an asymmetric 3D oscillator model where we have chosen the $z-$axis as the axis of deformation. Here $m$ is the nucleon mass, $\omega_z$ and $\omega_{\perp}$ are the nucleon oscillation frequencies along the $z-$axis and the perpendicular  plane respectively.  The deformation of the nucleus has the effect of splitting the degeneracy of the nuclear energy levels within the oscillator shell. The greater the deformation the greater the splitting. Each nucleon can be characterised by the set of quantum numbers $\left[N n_z \Lambda \pm\Omega\right]$ \cite{BohrMottVol2}  where $N = n_z + n_x + n_y$ and each $n_i$ ($i = x, y, z$) is the principal quantum number in the direction of $i$, $\Lambda$ and $\Omega$ are the nucleon's orbital and total angular momentum projection on the deformation axis respectivly. Each energy level is doubly degenerate for $\pm\Omega$. The average oscillator frequency   is given by $\hbar\bar{\omega} = \hbar/3\left(2\omega_{\perp} + \omega_z\right) \approx 45A^{-1/3} - 25A^{-2/3} \ \text{MeV}$\cite{Brown2016}. Second order tensor properties  in deformed nuclei exhibit a collective enhancement compared to vector properties. We indirectly include spin-orbit and angular momentum terms of the Nilsson Hamiltonian through the split level Nilsson energy plots in ref. \cite{BohrMottVol2}.

Consider the magnetic dipole moment  which is proportional to  the projection of the total angular momentum $\Omega$ to the nuclear axis. In an even-even nucleus all the nucleons are paired and therefore the magnetic dipole moment of the $+\Omega$ state cancels with the $-\Omega$ state resulting in no net magnetic dipole moment of the nucleus. For an odd $A$ nucleus  the magnetic dipole moment of the entire nucleus is simply the dipole moment of the odd nucleon. This is the well known Schmidt model of the nucleus. The cases for second order tensor properties of the nucleus are different. For  nucleons in the  $+\Omega$ and the $-\Omega$ states tensor properties  are additive. Therefore there is a collective effect and many nucleons contribute to the tensor properties of the nucleus.

Due to the rotation of the nucleus the tensor properties transform between the body-fixed (intrinsic) frame which rotates  and the laboratory frame. A second order tensor observable $T$ between these two frames has the relationship\cite{BohrMottVol2}
\begin{align} \label{eq:RotationalFactor}
T^{Lab} = \dfrac{I\left(2I - 1\right)}{\left(I + 1 \right)\left(2I + 3\right)}T^{Intrinsic},
\end{align}
where $I=I_z= \left|\Omega\right|$ is the projection of total nuclear angular momentum (nuclear spin) on the symmetry axis. This expression shows that only in nuclei with spin $I > 1/2$ we can detect these second order tensor properties.

\subsection{Quadrupole moment of neutron distribution in deformed nuclei}

The quadrupole moment tensor along the symmetry axis of the nucleus in cartesian coordinates is given by
\begin{align}\label{eq:QMomentStandard}
Q_{zz} = Q = 2\left<z^2\right> - \left<x^2\right> - \left<y^2\right> .
\end{align}
From the virial theorem for bound particle in a harmonic oscillator potential the average kinetic energy $\left<T\right>$ and average potential energy $\left<U\right>$ are equal. Using the well known energy spectrum for the harmonic oscillator $E_{n} = \hbar\omega(n + 1/2)$  the average of the square of the position in the $z$ direction is,
\begin{align}\label{eq:zExpectation}
m\omega_z^2\left<z^2\right>&= \hbar\omega_z\left(n_z + 1/2\right) 
\end{align}
  with similar relations for $\left<x^2\right>$ and $\left<y^2\right>$. Using equations (\ref{eq:QMomentStandard}), (\ref{eq:zExpectation}) and $n_x + n_y = N - n_z$ the contribution to the quadrupole moment from a single nucleon in the quantum state $\left[N n_z \Lambda \Omega\right]$ is given by
\begin{align} \label{eq:QMomentNucleon}
q_{i,\nu} &= \dfrac{\hbar}{m}\left[\dfrac{\left(2n_z + 1\right)_{i,\nu}}{\omega_z} - \dfrac{\left(N -n_z + 1\right)_{i,\nu}}{\omega_{\perp}}\right] ,
\end{align}
where $\nu = p,n$ for the $i$th proton or neutron respectively. The total quadrupole moment is the sum of all the respective nucleon quadrupole moment contributions
\begin{align}
Q_{\nu} &= \sum_{i} q_{i,\nu} \nonumber\\
&= \dfrac{\hbar}{m}\left[\dfrac{1}{\omega_z}\sum_{i}\left(2n_z + 1\right)_{i,\nu} - \dfrac{1}{\omega_{\perp}}\sum_{i}\left(N -n_z + 1\right)_{i,\nu}\right]. \label{eq:CollectiveMDim} 
\end{align}
\subsection{Lorentz invariance violation  in deformed nuclei}
Taking the expectation value of (\ref{eq:HamilQuadShift}) we find that the LLIV energy shift  for a nucleus with $N_{\nu}$ nucleons is given by 
\begin{align} \label{eq:MomentumEnergyShift}
<\delta H>_{\nu} = \dfrac{1}{6m}C_{0, \nu}^{(2)}\sum_{N=1}^{N_{\nu}}\left<I,I\left|\hat{M}\right|I,I\right> 
\end{align}
where the $\hat{M} = 2\hat{p}_z^2 - \hat{p}_x^2 - \hat{p}_y^2$ is the momentum tensor operator for $p_ip_j$ in the SME in cartesian coordinates. We use the standard notion $C_{0, \nu}^{(2)} = c_{xx} + c_{yy} -2c_{zz}$ \cite{Kostelecky1999}. We define the single nucleon LLIV momentum tensor as,
\begin{align} \label{eq:LLIVMomentum}
\bar{m}_{i, \nu} =  \left<I,I\left|2\hat{p}_z^2 - \hat{p}_x^2 - \hat{p}_y^2\right|I,I\right>.
\end{align}
Using the virial theorem for a harmonic oscillator ($\left<T\right>=\left<U\right>$) and energy spectrum again we have the average square of the momentum given by
\begin{align} 
\dfrac{\left<p_z^2\right>}{2m} &= \dfrac{\hbar\omega_z\left(n_z + 1/2\right)}{2} \label{eq:pExpectation}
\end{align}
with similar expressions for $x$ and $y$ coordinates. \\
\linebreak
Using equations (\ref{eq:LLIVMomentum}) and (\ref{eq:pExpectation}) we write the contribution to the LLIV tensor of a nucleon as,
\begin{align} \label{eq:MDimensionful}
\bar{m}_{i,\nu} &= \hbar m\left[\left(2n_z + 1\right)_{i,\nu}\omega_z - \left(N - n_z + 1\right)_{i,\nu}\omega_{\perp}\right].
\end{align}
The total LLIV tensor, $M_{\nu}$,  for the nucleus is the sum of all respective nucleons
\begin{align}
M_{\nu} &= \sum_{i,\nu} \bar{m}_{i,\nu} \nonumber \\
&= \hbar m\left[\omega_z\sum_{i}\left(2n_z + 1\right)_{i,\nu} - \omega_{\perp}\sum_{i}\left(N - n_z + 1\right)_{i,\nu}\right]. \label{eq:CollectiveQDim}
\end{align}
This will result in a quadrupole energy shift given by
\begin{align} \label{eq:LLIVEnergyShift}
\left<\delta H\right>_{\nu} = \dfrac{M_{\nu}}{6m} C_{0,\nu}^{(2)}.
\end{align}

\subsection{Example calculation of $Q_{\nu}$ and $M_{\nu}$ for the $^{173}$Yb nucleus}
Rewriting equations (\ref{eq:CollectiveMDim}), (\ref{eq:CollectiveQDim})  and, the relation between the longitudinal  and perpendicular frequencies in dimentionless quantities $\eta = \bar{\omega}/\omega_z$ and $\xi = \bar{\omega}/\omega_{\perp}$ we have the equations,
\begin{align}
3 &= \dfrac{1}{\eta} + \dfrac{2}{\xi} \label{eq:average} 
\end{align}
\begin{align}
Q_{\nu} = \dfrac{41.5\text{ MeV}\text{fm}^2}{\hbar\bar{\omega}}\left[\eta\sum_{i}\left(2n_z + 1\right)_{i,\nu} \right. \quad & \nonumber \\
 \left. - \xi\sum_{i}\left(N - n_z + 1\right)_{i,\nu}\right]&
\label{eq:QDimensionless}
\end{align}
\begin{align}
M_{\nu} = \hbar\bar{\omega}m\left[\dfrac{1}{\eta}\sum_{i}\left(2n_z + 1\right)_{i,\nu} \right. \quad & \nonumber \\ 
\left. - \dfrac{1}{\xi}\sum_{i}\left(N - n_z + 1\right)_{i,\nu}\right]. & \label{eq:MDimensionless}
\end{align}
The nucleon angular momentum dependence shows up in the order of the split energy branches in the Nilsson plots \cite{Nilsson1955, BohrMottVol2}.
The parameters $\eta$ and $\xi$ can be viewed as deformation parameters of the nuclei. For positive deformation $\eta > 1$ (prolate) and for negative deformation $\eta < 1$ (oblate). As there is a predominance on nuclear prolate deformations in nature most of the nuclei we consider will be prolate. \\

To illustrate the process of calculating NQMN and LLIV values we will present the calculation of $^{173}$Yb as an example. To begin we find the nuclear configuration of protons and neutrons in the deformed field using the energy level plots presented in \cite{Nilsson1955,BohrMottVol2} by filling each non-degenerate $\Omega$ energy branch with two nucleons until all the nucleons have been distributed. To find the correct deformation, $\delta$, from which to fill we use the experimental value of the nuclear spin (which is solely due to the unpaired nucleon). The filling must be done such that the unpaired nucleon is in the correct spin state. The nuclear spin of $^{173}$Yb is $5/2^{-}$ due to an unpaired neutron. Using the method described above we find that this is possible with a minimum deformation of $\delta \approx 0.3$. Here we will only present the configuration of the partially filled $N$ shells. The numbers for each full shell is given in Table \ref{table:FullShellNz}. The incomplete neutron and proton shells in $^{173}$Yb are $N = 5,6$ and $N = 4,5$ respectively. The nucleon configuration of the incomplete shells are given in Table \ref{table:YbConfig}.
\begin{table} 
\begin{tabular}{|c|c|}
\hline
Filled $N$ Shell & $\sum 2n_z + 1 = \sum N - n_z + 1$ \\ [2pt]
\hline
0 & 2\\
1 & 10\\
2 & 28\\
3 & 60\\
4 & 110\\
5 & 182 \\
\hline
\end{tabular}
\caption{This shows $\sum 2n_z + 1$ and $\sum N - n_z + 1$  for each completely filled $N$ shell in the Nilsson model. \label{table:FullShellNz} 
}
\end{table}
\begin{table*} 
\begin{tabular}{|p{1.5cm}|l|c|c|p{1.5cm}|l|c|c|}
\hline 
\multicolumn{2}{|c|}{\parbox[c][][c]{2cm}{\vspace{3pt}Neutrons in incomplete shells\vspace{3pt}}} & $\sum \left(2n_z + 1\right)_n$ & $\sum \left(N - n_z + 1\right)_n$ & \multicolumn{2}{|c|}{\parbox[c][][c]{2cm}{\vspace{3pt}Protons in incomplete shells\vspace{3pt}}} & $\sum \left(2n_z + 1\right)_p$ & $\sum \left(N - n_z + 1\right)_p$\\ 
\hline
 $2f_{7/2}$: &\parbox[c][][l]{2cm}{\begin{flushleft}
 $\pm 1/2[530]$, $\pm 3/2[521]$, $ \pm 5/2[512]$\end{flushleft} }&  30 & 24 & $1g_{7/2}$: & \parbox[c][][c]{2cm}{\begin{flushleft}
$\pm 1/2[431]$, $\pm 3/2[422]$, $ \pm 5/2[413]$
\end{flushleft} }& 30 & 18 \\
 $1h_{9/2}$: &  \parbox[c][][c]{2cm}{\begin{flushleft}
 $\pm 1/2[541]$, $\pm 3/2[532]$, $ 5/2[523]$\end{flushleft} } & 37 & 14 & $2d_{5/2}$: & \parbox[c][][c]{2cm}{\begin{flushleft}
 $\pm 1/2[420]$, $\pm 3/2[411]$ \end{flushleft} } & 16 & 14 \\
 $3p_{3/2}$: & \parbox[c][][l]{2cm}{\begin{flushleft}
$\pm 1/2[521]$\end{flushleft}} & 10 & 8 & $2d_{3/2}$: & \parbox[c][][c]{2cm}{ \begin{flushleft}
$\pm 1/2[521]$ \end{flushleft} } & 6 & 8 \\
 $1h_{11/2}$: & \parbox[c][][c]{2cm}{\begin{flushleft}
 $\pm 1/2[550]$, $\pm 3/2[541]$, $\pm 5/2[532]$, $\pm 7/2[523]$, $\pm 9/2[514]$, $\pm 11/2 [505]$\end{flushleft}}   & 72 & 42 & $1g_{9/2}$: &  \parbox[c][][c]{2cm}{\begin{flushleft}
$\pm 1/2[440]$, $\pm 3/2[431]$, $ \pm 5/2[422]$, $\pm 7/2[413]$, $\pm 9/2[404]$ \end{flushleft} }  & 50 & 30 \\
 $1i_{13/2}$: & \parbox[c][][l]{2cm}{\begin{flushleft}
 $\pm 1/2[660]$, $\pm 3/2[651]$, $\pm 5/2[642] $, $\pm 7/2[633]$\end{flushleft}} & 80 & 20 & $1h_{11/2}$: & \parbox[c][][l]{2cm}{\begin{flushleft}
$\pm 1/2[550]$, $\pm 3/2[541]$, $\pm 5/2[532]$, $\pm 7/2[523]$ \end{flushleft} } & 64 & 20 \\ \cline{1-2} \cline{5-6}
\multicolumn{2}{|c|}{Total} & 229 & 108 & \multicolumn{2}{c|}{Total}& 166 & 90  \\
\hline
\end{tabular}
\caption{Nilsson configuration of $^{173}$Yb nucleons: This table shows the nuclear configuration of nucleons in the Nilsson model for $^{173}$Yb generated from the Nilsson plots in \cite{BohrMottVol2}. This table shows only partially filled $N$ shells. All preceding shells are completely filled. \label{table:YbConfig}}
\end{table*}
Summing up the contribution for all the $^{173}$Yb filled shells from Table \ref{table:FullShellNz} and the partially filled shells from Table \ref{table:YbConfig} the total values for  neutrons are $\sum(2n_z + 1)_{n} = 439$, $\sum(N - n_z + 1)_n = 318$ and protons $\sum(2n_z + 1)_{p} = 266$, $\sum(N - n_z + 1)_p = 190$. Using (\ref{eq:RotationalFactor}) to transfer the measured value of the quadrupole moment in the lab frame $Q_p^{Lab} = 4.39$ barn (where 1 barn = 100 fm$^2$ = $10^{-24}$ cm$^{2}$) to the internal (rotating) frame we have $Q_{p}^{int} = 12.66\ \text{barn}$. Using equations (\ref{eq:QDimensionless}) and the values for $\sum\left(2n_z + 1\right)_{p}$, $\sum\left(N - n_z + 1\right)_p$ and $Q^{int}_p$ we find the dimensionless deformation parameters $\eta = 1.18$ and $\xi = 0.93$ for $^{173}$Yb. We can find the NQMN and the proton and neutron LLIV tensors using (\ref{eq:QDimensionless}) and (\ref{eq:MDimensionless}). Finally in the lab frame using (\ref{eq:RotationalFactor}) we have,
\begin{align*}
Q_{n}^{Lab} &= 4.53 \text{ barn} \\
M_{p}^{Lab} &= 54.30 \quad m\text{ MeV}\\
M_{n}^{Lab} &= 77.25 \quad m\text{ MeV}
\end{align*}
From (\ref{eq:LLIVEnergyShift}) we find that the LLIV energy shift of the nucleus is,
\begin{align*}
\left<\delta H\right>_{p} &= 9.05 \quad C^{(2)}_{0,p} \text{MeV} \\
\left<\delta H\right>_n &=  12.875 \quad C^{(2)}_{0,n} \text{MeV}
\end{align*}
Considering only the contribution of the unpaired neutron in the Schmidt model (see Section \ref{sec:Minimal} or Refs. \cite{Kostelecky1999,Flambaum2016}) gives energy shifts $\left<\delta H\right>_{p} = 0 \ C^{(2)}_{0,p} \text{MeV}$ and $\left<\delta H\right>_n = 0.8 \ C^{(2)}_{0,n} \text{MeV}$. The collective contribution of paired nucleons in the core gives non zero LLIV energy shifts for both protons and neutrons (in the Schmidt model either the proton or neutron LLIV shift will always be zero) and enhances the LLIV energy shifts by an order of magnitude.\\
This method can be completed with all the other deformed nuclei, the Nilsson quantum numbers can be found in Table \ref{table:NzNumbers} and the quadrupole moment values are presented in Table \ref{table:LLIVDeformed}.\\
\begin{table}[h!]
\center
\begin{tabular}{|c|c|c|c|c|c|}
\hline
 & \multicolumn{2}{|c|}{Proton ($\Sigma_p$)} & \multicolumn{2}{|c|}{Neutron ($\Sigma_n$)} & \multirow{2}{*}{$\dfrac{\bar{\omega}}{\omega_z}$} \\
\cline{2-5}
 & $2n_z + 1$ & $N - n_z + 1$ & $2n_z + 1$ & $N - n_z + 1$ &  \\
\hline
$^{9}$Be   & 8   & 4   & 9   & 6   & 1.68\\
$^{21}$Ne  & 22  & 14  & 25  & 16  & 1.30\\
$^{27}$Al  & 29  & 21  & 30  & 24  & 1.12\\
$^{151}$Eu & 199 & 177 & 312 & 276 & 1.08\\
$^{153}$Eu & 241 & 162 & 350 & 272 & 1.11\\
$^{163}$Dy & 250 & 174 & 389 & 287 & 1.12\\
$^{167}$Er & 260 & 182 & 417 & 302 & 1.17\\
$^{173}$Yb & 266 & 190 & 439 & 318 & 1.18\\
$^{177}$Hf & 264 & 202 & 439 & 331 & 1.18\\
$^{179}$Hf & 264 & 202 & 447 & 341 & 1.16\\
$^{181}$Ta & 285 & 199 & 468 & 338 & 1.08\\
$^{229}$Th & 362 & 268 & 625 & 502 & 1.22\\
\hline
\end{tabular}
\caption{Sum of proton and neutron Nilsson quantum numbers and deformation parameters for deformed nuclei: This table shows the sum of $2n_z + 1$ and $N - n_z + 1$ for all nucleons in the nuclei.}
\label{table:NzNumbers}
\end{table}
To understand the propagation of error in the calculations consider an error of 5\% in $Q_p$ (experimental value is $Q_p = 2.80(4)$ barn). Using (\ref{eq:CollectiveMDim}), (\ref{eq:CollectiveQDim}) with the Nilsson numbers for $^{173}$Yb results in an error of $\approx 5\%$ for  $Q_n$  and an error of $\approx 25\%$ for $M_{\nu}$. This large error is not unique to our calculation of $M_{\nu}$ since it involves the subtraction of large numbers. Consider the results from \cite{Brown2016} where a sophisticated $s-d$ model resulted in similar uncertainty for slight variations of an effective charge for the quadrupole moment operator.

\section{Nuclei with a small deformation} \label{sec:Minimal}

For nuclei with very small deformations the splitting of energy levels with different angular momentum projections is small.
 In these circumstances the effect of nuclear pairing becomes significant resulting in mixing of the nucleon configurations. Therefore the Nilsson model approach is no longer applicable as it assumes that there is no mixing when counting the nucleon occupation numbers. For these near-spherical nuclei we need to approach the second order tensor properties differently.  It is well known that the Schmidt model (valence, $Q_{\nu, val}$) value of nuclear quadrupole is smaller than the true quadrupole value. Therefore we assume that this discrepancy is explained by the quadrupole moment due to a small deformation (deformed, $Q_{\nu, def}$), i.e. the true quadrupole moment is the sum of these two contributions, $Q_{\nu} = Q_{\nu, val} + Q_{\nu, def}$. We also assume that $Q_{\nu,deformed}$ for protons and neutrons is related to the total number of protons and neutrons,
\begin{align} \label{eq:Val+Deformed}
\dfrac{1}{Z}Q_{p, def} = \dfrac{1}{N}Q_{n,def}.
\end{align}
Then using measured $Q_{p}$ values from \cite{Stone2005} we can find an estimate for the neutron quadrupole moment of slightly deformed nuclei. As an example consider the $^{201}$Hg nucleus which has an electric quadrupole moment $Q_{p} = 0.35 $ barn with a valence neutron. There is no proton Schmidt contribution to $Q_{p}$ meaning the quadrupole moment due to the deformation is 
\begin{align*}
Q_{p,def} = Q_{p} = 0.35 \ \text{barn}
\end{align*}
From Eq. (\ref{eq:Val+Deformed}) the contribution to $Q_{n}$ due to deformation is $Q_{n,def} = 0.53 \ \text{barn}$. The Schmidt model contribution of a valence nucleon is given by \cite{BohrMottVol1, Flambaum2016}
\begin{align} \label{eq:SchmidtQuad}
Q_{n}^{Lab} &= -\dfrac{I - 1/2}{I + 1}\left<r^2\right> \\
&= -\dfrac{I - 1/2}{I + 1}0.009A^{2/3} \text{ barn}.
\end{align}
 The $^{201}$Hg nucleus has a valence neutron in the $f_{5/2}$ state with angular projection $I =3/2$. Using \ref{eq:SchmidtQuad} the valence contribution is $Q_{n,val} = -0.15 \ \text{barn}$. Therefore the total NQMN for $^{201}$Hg is $Q_{n} = 0.38 \ \text{barn}$ where all values are in the labratory frame. As expected this value is larger than $Q_{p}$. For the LLIV tensor we use the method outlined in \cite{Flambaum2016} which relates the LLIV energy shift to the quadrupole moment of the nucleus,
\begin{align} \label{eq:LLIVsd}
\left<\delta H\right>_{\nu} = \dfrac{M_\nu}{6m} C_{0, \nu}^{(2)} = 1100A^{-2/3}Q_{\nu} C_{0, \nu}^{(2)} \ \text{MeV}.
\end{align}
In this work we use the total quadrupole moment including the valence and deformed contribution discussed above. Similar to the NQMN this will give non zero values for both nucleons unlike the Schmidt model. Using \ref{eq:LLIVsd} for $^{201}$Hg nucleus we have $\left<\delta H\right>_{n} = 17 \ C_{0, n}^{(2)} \ \text{MeV}$ and $\left<\delta H\right>_{p} = 11.2 \ C_{0, p}^{(2)} \ \text{MeV}$. Similar calculations can be performed for $Q_n$ and $M_{\nu}$ in other slightly deformed nuclei such as $^{131}$Xe, $^{133}$Cs which are presented in Table \ref{table:LLIVDeformed}. In \cite{Brown2016} the LLIV and quadrupole moments were calculated for $^{21}$Ne, $^{131}$Xe and $^{201}$Hg numerically using a self-consistent mean field theory. 
Our results for $^{21}$Ne, where we  use the large deformation method, are in a reasonable agreement  with Ref.  \cite{Brown2016} results for both $Q_{n}$ and $M_{\nu}$ ($Q_n = 0.097 \ \text{barn}$, $M_{p} = 2.8 \ m\text{ MeV}$ and $M_n = 4.2 \ m\text{ MeV}$). For nuclei  $^{201}$Hg  and $^{131}$Xe, where we use the small deformation method, there is a reasonable agreement for $Q_n$ and significant differences for $M_{\nu}$. In Ref.  \cite{Brown2016}  they obtained for $^{201}$Hg  $Q_n = 0.584 \ \text{barn}$, $M_{p} = -20.5 \ m\text{ MeV}$ and $M_n = 1.5 \ m\text{ MeV}$,  and for $^{131}$Xe their results are $Q_n = -0.136 \ \text{barn}$, $M_{p} = -4.7 \ m\text{ MeV}$ and $M_n = 5.17 \ m\text{ MeV}$.

\begin{table*}[t!]
\center
\begin{tabular}{|c|c|c|c|c|c|c|c|c|}
\hline
Nuclei & $I_t$ & $Q_p$ (barn) & $Q_n$ (barn) &$M_p$ ($m$ MeV)& $M_n$ ($m$ MeV)& \parbox[c][][c]{3cm}{ \vspace{3pt}$\dfrac{<\delta H>_p}{C_{0,p}^{(2)}}$ (MeV) \vspace{2pt}}&  $\dfrac{<\delta H>_n}{C_{0,n}^{(2)}}$ (MeV) & $Q_{W}^{(2)}$ (barn)\\
\hline
$^{9}$Be & $\tfrac{3}{2}^{-}$ & +0.0529(4)& +0.053 & -0.14 & -5.88 & -0.024 & -1 & -0.05\\
$^{21}$Ne & $\tfrac{3}{2}^+$ & +0.103(8) & 0.12 & 3.19 & 3.36 & 0.53 & 0.56 & -0.11\\
$^{27}$Al & $\tfrac{5}{2}^+$ & +0.150(6)  & 0.129 & 17.12 & 7.26 & 2.85 & 1.21 & -0.12\\
$^{131}$Xe* & $\tfrac{3}{2}^+$ & -0.114& -0.070 & -29.4 & -18 & -4.9 & -3 & -0.009\\
$^{133}$Cs* & $\tfrac{7}{2}^+$ & -0.00355 & 0.2 & -0.9 & 53 & -0.15 & 9 & -0.20\\
$^{151}$Eu & $\tfrac{5}{2}^+$ & +0.87(2)& 1.4 & 2.45 & 8 & 0.41 & 1.3 & -1.33\\
$^{153}$Eu & $\tfrac{5}{2}^+$ & +2.28(9)& 2.53 & 127 & 81 & 21 & 13.5 & -2.35\\ 
$^{163}$Dy & $\tfrac{5}{2}^-$ & +2.318(2) & 3.3 & 103.4 & 115.5 & 17 & 19 & -3.10\\
$^{167}$Er & $\tfrac{7}{2}^+$ & 3.57(3)& 5.47 & 90 & 107 & 15 & 18 & -5.17\\
$^{173}$Yb & $\tfrac{5}{2}^-$ & +2.8(4) & 4.5 & 54 & 77 & 9 & 13 & -4.26\\
$^{177}$Hf & $\tfrac{7}{2}^-$ & +3.37(3) & 5.73 & 16.35 & 45  & 2.7 & 7.5 & -5.44\\
$^{179}$Hf & $\tfrac{9}{2}^+$ & +3.79(3)& 6.5 & 35.33 & 64 & 6 & 10.5 & -6.17\\
$^{181}$Ta & $\tfrac{7}{2}^+$ & +3.17(2) & 5.33 &  188 & 269.5 & 31 & 45 & -5.05\\
$^{201}$Hg* & $\tfrac{3}{2}^+$ & +0.35 & 0.53 & 67.2 & 102 & 11.2 & 17 & -0.50\\
$^{229}$Th & $\tfrac{5}{2}^+$ & +4.3(9) & 6.62 & 14 & -78 & 2 & -13 & -6.25\\
\hline
\end{tabular}
\caption{Results for LLIV and quadrupole tensors for different deformed nuclei: (All used $Q_p$ values have been compiled in \cite{Stone2005}.)  This table shows the proton ($Q_p$) and neutron ($Q_n$) electric quadrupole moments, energy shifts due to Lorentz violation ($\left<\delta H \right>_{\nu}$) and the weak quadrupole moments ($Q_{W}^{(2)}$) in $^{9}$Be, $^{21}$Ne, $^{27}$Al, $^{131}$Xe, $^{133}$Cs,  $^{163}$Dy, $^{173}$Yb, $^{177}$Hf, $^{179}$Hf, $^{181}$Ta, $^{201}$Hg and $^{229}$Th. All quantities are in the lab frame. Nuclei marked with an asterix (*) are near spherical nuclei. \label{table:LLIVDeformed}}

\end{table*}

\section{Weak quadrupole moments and parity nonconservation in atomic and molecular systems\label{sec:PNC}} 
As previously mentioned above a consequence of studying the NQMN will be a further insight into parity nonconservation (PNC) effects in atomic and molecular systems. PNC has been intensively studied in atomic systems (see e.g. book \cite{KhriplovichPNC} and reviews \cite{RobertsReview, GingesReview}).  The $P$-odd weak nucleon-electron interaction is given by,
\begin{align} \label{eq:eNWeak}
H_W = -\frac{G_F}{2\sqrt{2}}\gamma_{5}\left[Zq_{w, p}\rho_{p}(r) + Nq_{w, n}\rho_{n}(r)\right]
\end{align}
Here $G_F$ is the Fermi weak constant, $q_{w,\nu}$ and $\rho_{\nu}(r)$ are the nucleon weak charge  and density of protons or neutrons  normalised to 1. It is well known that the magnitude of the neutron weak charge is significantly larger than that of the proton. Not including radiative corrections the weak charges  are given by 
\begin{align*}
q_{w,p} &= 1 - 4\sin^2\theta_W \approx 0.08 ,\\
q_{w,n} &= -1,
\end{align*}
where $\theta_W$ is the Weinberg angle.  Previously the interaction (\ref{eq:eNWeak}) treated either the shapes of the proton and neutron densities  the same  or included a correction due to some neutron skin (see e.g. review \cite{RobertsReview}).  The quadrupole moment in the nuclear density produces the tensor weak interaction which is proportional to the weak quadrupole moment (WQM) defined as  \cite{FDC17}
\begin{align*}
Q_{W}^{(2)} = q_{w,p}Q_{p} + q_{w,n}Q_n.
\end{align*}
Similar to the weak charge of a nucleus the WQM is dominated by the neutron contribution:  $Q_{W}^{(2)} \approx q_{w,n}Q_n$  with a small correction due to the proton contribution. \\

The nuclear WQM induces PNC effects in atomic and molecular systems where the effective single electron  PNC Hamiltonian for the nuclear WQM in atomic systems is presented in ref. \cite{FDC17} and is given by,
\begin{align*}
H_{WQM}=-\dfrac{G_F}{2\sqrt{2}}\gamma_5Y_{20}\rho_{0}\dfrac{\sqrt{5\pi}Q_{W}^{(2)}}{\left<r^2\right>}
\end{align*}
where $G_F$ is the Fermi weak constant, $\gamma_5$ is the standard Dirac matrix, $Y_{20}$ is the spherical harmonic, $\rho_0$ is the spherical nucleon density and $\left<r^2\right>$ is the mean squared nuclear radius. While calculations of these PNC effects is outside the scope of this work they can be observed in atomic and molecular systems in many ways (see refs. \cite{FDC17, RobertsReview, KhriplovichPNC, GingesReview}) using interference of forbidden electric dipole transition amplitude with M1 ( or E2) amplitude between the states of equal parity \cite{FDC17}.

Also the PNC effects of the tensor weak interaction which has different selection rules are strongly enhanced  and can be measured in atoms and molecules having close opposite parity energy levels with the difference of the electron angular momenta equal to 2. Corresponding states can be mixed by the tensor weak interaction but not the scalar (proportional to the  weak charge) and vector (proportional to the nuclear anapole moment) components. If the difference of the electron angular momenta is 1 the effects of the anapole and the weak quadrupole may be separated due the difference in their contributions to the different hyperfine components of the electromagnetic transitions. Corresponding atomic calculations have been performed in Ref.  \cite{FDC17}. Measurements of the tensor PNC effects in atoms and molecules will allow one to extract  the neutron quadrupole moment of the nuclei. \\

We present the nuclear weak quadrupole moments  $Q_{W}^{(2)}$ in Table \ref{table:LLIVDeformed}. As expected heavily deformed nuclei with a large number of neutrons have large WQMs.

\section{Conclusion}
We presented a versatile semi-empirical method to calculate several second order tensor properties of nuclei which are enhanced in deformed nuclei including the weak quadrupole moment. Though we present values for only a few deformed nuclei the method can be applied to many nuclei of interest. In particular the highly deformed $^{181}$Ta, $^{167}$Er, $^{163}$Dy and $^{153}$Eu have large enhancement of tensor properties and are particularly promising candidates for further study. The theoretical results presented should facilitate experimentalists in probing the nuclear structure and fundamental physics, specifically the previously unstudied quadrupole distribution of neutrons and the violations of Lorentz symmetry. Currently such systems are of high interest  as they allow the study of physics beyond the standard model with low energy systems. \\

This enhancement in deformed nuclei can also be applied for other second order tensor observables of interest such as the time-reversal and parity violating magnetic quadrupole moment. Further study of these tensor properties in deformed nuclei  could lead to further understanding of fundamental physics.\\

 This work is supported in part by the Australian Research Council and the Gutenberg Fellowship. 

\bibliographystyle{apsrev4-1}
\bibliography{Deformed_Nuclei}
\end{document}